\documentclass{svjour3}
\journalname{Theoretical Chemistry Accounts}
\usepackage{amssymb}
\usepackage{gensymb}
\usepackage{graphicx}

\renewcommand{\vec}[1]{{\rm\bf #1}}
\newcommand{\de}{{\rm d}}
\newcommand{\frct}[2]{{\textstyle\frac{#1}{#2}}}
\newcommand{\ka}{\kappa}
\newcommand{\mt}{\mathfrak{t}}
\newcommand{\mr}{\mathfrak{r}}
\newcommand{\mf}{\mathfrak{f}}
\newcommand{\mm}{\mathfrak{m}}

\begin{document}

\title{A reversible numerical integrator \\ of the isokinetic equations of motion}
\author{Dimitri N. Laikov}
\institute{D. N. Laikov \at
 Chemistry Department, Moscow State University, 119991 Moscow, Russia \\
\email{laikov@rad.chem.msu.ru}
}

\date{Received: date / Accepted: date}

\maketitle

\begin{abstract}
An explicit second-order numerical method
to integrate the isokinetic equations of motion
is derived by fitting circular arcs
through every three consecutive points
of the discretized trajectory,
so that the tangent and the curvature
satisfy the equations exactly at every central point.
This scheme is reversible and robust,
and allows an adaptive step size control.
Its performance is tested by computing
the thermodynamic properties of simple pair-potential models,
and its chemical application is shown
for the global search for stable structures,
using canonical sampling and energy minimization,
of hydrogen-bonded molecular clusters.
\end{abstract}

\section{Introduction}

Early in the history of computers, numerical methods
for the atomistic simulation of thermodynamic properties of matter
began their evolution~\cite{MRRTT53,AW59}.
The idea of the isokinetic equations of motion
seems to show up first quietly in a work~\cite{W72} on liquid salts,
where a numerical scheme~\cite{V67} to solve the Newtonian equations
was modified, by what was later called ``velocity rescaling'',
to keep the kinetic energy constant,
the underlying differential equation,
although not written out at that time,
was later rediscovered~\cite{HLM82,EM83}
and its properties were carefully studied~\cite{DM96}.
Optimal numerical integration of this kind of equations
seems to have drawn only limited attention
(which can be due in part to the shift
to other deterministic thermostats~\cite{N84a,N84b,H85}),
the original scheme~\cite{W72} lacks reversibility,
the integrators~\cite{Z97,MMT03} based on operator-splitting~\cite{MTTK96}
were developed that respect this condition.
Here we design an explicit reversible integrator
of the isokinetic equations of motion
based on the local circular arc interpolation ---
the simplest analytical curve
naturally parametrized by its length,
with the simplest (constant) curvature,
fits well to the isokinetic trajectory
and plays the same role as the parabola
in the Verlet~\cite{V67} method ---
it also shares some kinship
with the well-known implicit second-order
integrator~\cite{GS89,GS91,G71}
of the first-order differential equations
of the steepest descent path.

\section{Theory}
\label{seq:theory}

The isokinetic equations of motion for a system of $N$ atoms
can be written in the form
\begin{equation}
\label{eq:kin}
\ddot{\vec{r}}(t) =
\left(
1 - \frac{\dot{\vec{r}}(t)\dot{\vec{r}}(t)\cdot}{|\dot{\vec{r}}(t)|^2}
\right) \vec{f}\bigl(\vec{r}(t)\bigr)
\end{equation}
where $\vec{r}(t)$ is the $3N$-dimensional vector
of their cartesian coordinates as a function of time,
and $\dot{\vec{r}}(t)$ and $\ddot{\vec{r}}(t)$ mean its first
and second time derivatives.
We assume unit masses of all atoms for simplicity, moreover,
the masses play no role for sampling the canonical distribution
in the coordinate space.
The vector of forces
\begin{equation}
\label{eq:f}
\vec{f}(\vec{r}) = -\gamma \frac{\partial U(\vec{r})}{\partial\vec{r}}
\end{equation}
can be taken as a scaled gradient
of the potential energy function $U(\vec{r})$,
 with $1/\gamma = (3N - 1) T$ and $|\dot{\vec{r}}(t)| = 1$
now and in the following
we get a trajectory in the configuration space $\vec{r}(t)$
parametrized by its length (the now fictitious time $t$),
which, \textit{if ergodic}, would have
the density distribution $\sim\exp\bigl(-U(\vec{r})/T\bigr)$
in the limit $t\to\infty$, where the temperature $T$ is given in energy units.
To start the trajectory, the positions $\vec{r}(t_0)$ and
the velocities $\dot{\vec{r}}(t_0)$ at time $t_0$ are taken as input.

Numerical integration of Eq.~(\ref{eq:kin}) works by computing
an approximate solution $\{\vec{r}_n\} \equiv \{\vec{r}(t_n)\}$
at discrete points in time $\{t_n\}$,
the steps $t_{n+1} - t_n > 0$ being small enough
to get a good accuracy.
A local interpolation around each point
\begin{equation}
\vec{r}_n(t) = \vec{r}_n + \vec{c}_n (t - t_n)
\end{equation}
can be found that goes through the three consecutive points
$\vec{r}_{n-1}$, $\vec{r}_n$, and $\vec{r}_{n+1}$
at some $t$,
and making it satisfy Eq.~(\ref{eq:kin}) at the central point $\vec{r}_n$
leads to a (generally nonlinear) dependence between the three points,
so that $\vec{r}_{n+1}$ is defined by $\vec{r}_{n-1}$ and $\vec{r}_n$
or in the same way $\vec{r}_{n-1}$ by $\vec{r}_n$ and $\vec{r}_{n+1}$,
thus a \textit{reversible} trajectory can be computed.

In this work we put forward the circular arc interpolation
\begin{equation}
\label{eq:c}
\vec{c}_n(s) =
\frac{\sin(\ka_n s)}{\ka_n} \vec{z}_n
+ \frac{1 - \cos(\ka_n s)}{\ka_n^2}
 (1 - \vec{z}_n \vec{z}_n \cdot) \vec{f}_n
\end{equation}
with the scalar curvature
\begin{equation}
\label{eq:kan}
\ka_n = \bigl|(1 - \vec{z}_n \vec{z}_n \cdot) \vec{f}_n \bigr|
= \sqrt{|\vec{f}_n|^2 - (\vec{z}_n \cdot\vec{f}_n)^2}
\end{equation}
and the (unit) tangent vector $\vec{z}_n$.
By construction, it satisfies Eq.~(\ref{eq:kin})
at the central point for any unit $\vec{z}_n$,
but making it go through either of the two other points
\begin{eqnarray}
\vec{r}_{n-1} &=& \vec{r}_n + \vec{c}_n (-\bar{s}_n) \\
\label{eq:rn1}
\vec{r}_{n+1} &=& \vec{r}_n + \vec{c}_n (s_{n+1})
\end{eqnarray}
uniquely defines the tangent as either
\begin{equation}
\label{eq:zn}
\vec{z}_n =
\frac{\vec{x}_n + \frct12 |\vec{x}_n|^2 \,\vec{f}_n}
{\left|\vec{x}_n + \frct12 |\vec{x}_n|^2 \,\vec{f}_n\right|} ,
\end{equation}
\begin{equation}
\vec{x}_n \equiv \vec{r}_n - \vec{r}_{n-1} ,
\end{equation}
or likewise
\begin{equation}
\label{eq:zn1}
\vec{z}_n =
\frac{\vec{x}_{n+1} - \frct12 |\vec{x}_{n+1}|^2 \,\vec{f}_n}
{\left|\vec{x}_{n+1} - \frct12 |\vec{x}_{n+1}|^2 \,\vec{f}_n\right|} .
\end{equation}
A somewhat lengthy derivation of Eqs.~(\ref{eq:zn}) or~(\ref{eq:zn1})
is given in Appendix~\ref{sec:zeq}.

The local arc lengths $s_n \approx \bar{s}_n$
in general are not equal,
\begin{eqnarray}
\label{eq:sn1}
      s_n &=& \frac{\arcsin(\ka_{n-1}\;\vec{z}_{n-1} \cdot\vec{x}_n)}{\ka_{n-1}}, \\
\label{eq:sn}
\bar{s}_n &=& \frac{\arcsin(\ka_n\;\vec{z}_n \cdot\vec{x}_n)}{\ka_n},
\end{eqnarray}
so the total path length can be computed as
\begin{equation}
t_n = t_{n-1} + \frct12 (s_n + \bar{s}_n).
\end{equation}
For a statistical average of some property $P(\vec{r})$,
such as pressure,
\begin{equation}
\label{eq:pr}
\left< P \right> =
\frac{\sum\limits_n w_n P(\vec{r}_n)}
     {\sum\limits_n w_n}
\end{equation}
it seems natural to take the weights
\begin{equation}
w_n = \frct12 (\bar{s}_n + s_{n+1}) .
\end{equation}
With all this in mind,
the computation of the isokinetic trajectory
starts with
$ \vec{z}_0 = \dot{\vec{r}}(t_0)$
and makes steps forward through Eqs.~(\ref{eq:kan}), (\ref{eq:c}),
and (\ref{eq:rn1}), and also Eq.~(\ref{eq:zn}) for $n > 0$.

There is much freedom in the choice of arc lengths $s_n$,
the simplest would be to set $s_{n+1} = \bar{s}_n$,
then it can be shown that
\begin{equation}
\label{eq:xrf}
\vec{x}_{n+1} = \vec{c}_n(\bar{s}_n) = -(1 - 2\vec{z}_n \vec{z}_n \cdot)\vec{x}_n ,
\end{equation}
thus every next step is a negative reflexion of the step before
against the tangent of Eq.~(\ref{eq:zn}),
and as such it conserves the length $|\vec{x}_{n+1}| = |\vec{x}_n| = x$,
a procedure elegant in its simplicity.
Setting the step lengths to some values
\begin{equation}
|\vec{x}_{n+1}| = \bigl|\vec{c}_n(s_{n+1})\bigr| = x_{n+1}
\end{equation}
yields the arc lengths
\begin{equation}
s_{n+1} = \frac{2\arcsin\left(\frct12 \ka_n x_{n+1}\right)}{\ka_n} ,
\end{equation}
and this can be done at the first step followed by the use of Eq.~(\ref{eq:xrf}).

For similar systems, setting the step length $x = a\sqrt{N}$
may be meaningful,
assuming each atom to move by $\sim a$ on every step,
but hot atoms can show up, from time to time,
moving by up to $a\sqrt{N}$,
which can become dangerous for large $N$.
We think that a more mindful step size control would be
to limit the atomic moves
\begin{equation}
\label{eq:a}
\max\limits_i |\vec{x}_{i,n+1}| =
\max\limits_i \bigl|\vec{c}_{in}(s_{n+1})\bigr| = a,
\end{equation}
where $i$ labels the $i$-th atom's vector components,
and also to limit the bend angle
\begin{equation}
\label{eq:b}
\ka_n s_{n+1} \le \alpha,
\end{equation}
the solution being
\begin{equation}
\label{eq:sa}
s_{n+1} = \frac1{\ka_n}
\min \left(\min\limits_i \theta_{i,n+1}, \,\alpha \right)
\end{equation}
where the atomic values $\theta_{i,n+1}$ are the smallest roots
\begin{equation}
\label{eq:pin}
p\left(\theta_{i,n+1},
\frac{|\vec{z}_{in}|^2}{\ka^2_n a^2},
\frac{\vec{z}_{in}\cdot\vec{y}_{in}}{\ka^3_n a^2},
\frac{|\vec{y}_{in}|^2}{\ka^4_n a^2}
\right) = 0,
\end{equation}
\begin{equation}
\vec{y}_n \equiv (1 - \vec{z}_n \vec{z}_n \cdot) \vec{f}_n,
\end{equation}
of the functions
\begin{eqnarray}
\nonumber
p(\theta,u,w,v) &=&
u \sin^2 \theta + 2 w (1 - \cos \theta) \sin \theta \\
&+& v (1 - \cos \theta)^2 - 1,
\end{eqnarray}
or $\theta_{i,n+1} = \pi$ if there are no roots for some $i$,
the roots $\theta$ of these functions
are found through the solution of a quartic equation
as detailed in Appendix~\ref{sec:seq}.
Care should be taken if $\ka_n$ is small (our threshold is $\ka_n < 2^{-12}$),
then the trigonometric functions can be expanded in powers around zero,
and with $\theta_{i,n+1} = \ka_n s_{i,n+1}$ Eq.~(\ref{eq:pin})
is replaced by
\begin{equation}
\label{eq:qin}
q\left(s_{i,n+1},
\frac{|\vec{z}_{in}|^2}{a^2},
\frac{\vec{z}_{in}\cdot\vec{y}_{in}}{a^2},
 \frac{|\vec{y}_{in}|^2}{4a^2}
-\frac{\ka^2_n |\vec{z}_{in}|^2}{3a^2}
\right) = 0,
\end{equation}
the roots $s_{i,n+1}$ now being of the quartic function
\begin{equation}
q(s,u,w,v) = u s^2 + w s^3 + v s^4 - 1,
\end{equation}
finding them in a stable way needs some further tricks
as also explained at the end of Appendix~\ref{sec:seq}.

A simple bisection can also be used to solve Eqs.~(\ref{eq:a}) and~(\ref{eq:b}),
but may be too slow.
An approximation
\begin{equation}
s_{n+1} = 
\min \left( \frac{a}{\max\limits_i |\vec{z}_{in}|}, \,\frac{\alpha}{\ka_n} \right)
\end{equation}
could be used instead of Eq.~(\ref{eq:sa}), which does not respect Eq.~(\ref{eq:a})
exactly but does limit the atomic moves in some way,
although the reversibility would then be lost.

If the infinitesimal translations and rotations of the whole system
have to be removed from the trajectory
(or some other constraints enforced),
the tangents can be projected
\begin{equation}
\label{eq:o}
\vec{z}_n^\circ = \vec{O}_n \vec{z}_n
\end{equation}
by the well-known matrices $\vec{O}_n$ at each point,
and then not only $\vec{z}_n$ in Eq.~(\ref{eq:c})
should be replaced with $\vec{z}_n^\circ$,
but also
\begin{equation}
\vec{f}_n^\circ = 
-2\frac{\vec{x}_n}{|\vec{x}_n|^2}
\end{equation}
should be substituted for $\vec{f}_n$,
the latter follows from Eq.~(\ref{eq:fx}) in Appendix~\ref{sec:zeq}.

A retrospective error estimate
\begin{equation}
\label{eq:e}
\varepsilon_n = \left|\vec{r}_n(t_{n-2}) - \vec{r}_{n-2}\right|
\end{equation}
and its mean $\left<\varepsilon\right>$ and largest $\varepsilon_\mathrm{max}$ values
over the trajectory can be easily computed
and used to judge the local accuracy of the integrator.

Some workers in the field of molecular dynamics
may like to have the working equations with masses
at hand, we show in Appendix~\ref{sec:mass}
how a change of variables converts
our equations into such form,
and also compare
the velocity-rescaling~\cite{W72} integrator
with ours.

\section{Numerical tests}

We test our numerical integrator
on finite atomic or molecular systems
inside a spherical vessel of volume $V$,
adding a term
\begin{equation}
U_0(\vec{r}, V) = \sum\limits_i
u_0 \left( |\vec{r}_i| - \sqrt[3]{\frac{3V}{4\pi}} \right)
\end{equation}
to the potential energy,
with a particle-wall interaction
\begin{equation}
\label{eq:u0}
u_0(r) = \left\{
\begin{array}{rr}
0, & r \le 0 \\
\frct12 cbr^2/(b - r), & 0 < r < b \\
\infty, & r \ge b
\end{array}
\right. ,
\end{equation}
which allows the pressure to be computed as the average
of the volume derivative
\begin{equation}
P(\vec{r}) = -\frac{\partial U_0(\vec{r},V)}{\partial V}
\end{equation}
over the trajectory as in Eq.~(\ref{eq:pr}).
In the limit $b\to\infty$ Eq.~(\ref{eq:u0})
is simply a switched harmonic potential with force constant $c$,
a finite $b$ will keep all atoms from getting deeper than $b$
into the elastic wall.
For the ideal gas inside such a non-ideal vessel,
the partition function and the pressure
can be computed analytically for $b=\infty$,
thus the fugacity factor
\begin{equation}
\label{eq:p0}
\varphi = \frac{PV}{NT} =
\frac{1 + \sqrt{\pi}\mu + \mu^2}
     {1 + \frct32\sqrt{\pi}\mu + 3\mu^2 + \frct34\sqrt{\pi}\mu^3}
\end{equation}
with
\begin{equation}
\mu = \sqrt{\frac{2T}{c}}\cdot\sqrt[3]{\frac{4\pi}{3V}} .
\end{equation}
In the numerical computation of the isokinetic trajectory,
the projection of rotations (but not translations) of the whole system
in Eq.~(\ref{eq:o}) is helpful, and we do it throughout this work.

A model system of atoms with only pairwise interactions
\begin{equation}
U_2(\vec{r}) = \sum\limits_{i<j}
u\bigl(|\vec{r}_i - \vec{r}_j|\bigr)
\end{equation}
with either the purely repulsive potentail of Hertzian~\cite{H882} spheres
\begin{equation}
u(r) = \left\{
\begin{array}{rr}
(1 - r)^{5/2}, & r < 1 \\
0, & r \ge 1
\end{array}
\right.
\end{equation}
or the more realistic Lennard-Jones~\cite{J24} potential
\begin{equation}
u(r) = r^{-12} - 2r^{-6}
\end{equation}
should be good enough for testing the properties
of the numerical integrator.

\begin{table} 
\caption{Computed pressure $P$ of $N=13$ Hertzian spheres
at $T=8$ in a spherical vessel of volume $V$,
given as $\varphi=\frac{PV}{NT}$.}
\label{tab:h8}
\begin{tabular}{rrrrrr}
\hline\noalign{\smallskip}
$a$ &
 \multicolumn{1}{r}{$1/4$}  &
 \multicolumn{1}{r}{$1/8$}  &
 \multicolumn{1}{r}{$1/16$}  &
 \multicolumn{1}{r}{$1/8$} \\
$n,n_0$ &
 \multicolumn{1}{r}{$2^{30}, 2^{24}$} &
 \multicolumn{1}{r}{$2^{30}, 2^{24}$} &
 \multicolumn{1}{r}{$2^{30}, 2^{24}$} &
 \multicolumn{1}{r}{$2^{36}, 2^{25}$} &
 \\
\noalign{\smallskip}\hline\noalign{\smallskip}
$V/N$&\multicolumn{1}{c}{$\varphi$} &
 \multicolumn{1}{c}{$\varphi$} &
 \multicolumn{1}{c}{$\varphi$} &
 \multicolumn{1}{c}{$\varphi_\mathrm{walk}$} &
 \multicolumn{1}{r}{$\varphi_\mathrm{ideal}$} \\
\noalign{\smallskip}\hline\noalign{\smallskip}
   1 & 0.8607 & 0.8642 & 0.8645 & 0.8661 & 0.8598 \\
   2 & 0.8861 & 0.8890 & 0.8893 & 0.8906 & 0.8867 \\
   4 & 0.9076 & 0.9099 & 0.9102 & 0.9110 & 0.9088 \\
   8 & 0.9259 & 0.9275 & 0.9277 & 0.9281 & 0.9269 \\
  16 & 0.9411 & 0.9419 & 0.9422 & 0.9422 & 0.9415 \\
  32 & 0.9538 & 0.9538 & 0.9540 & 0.9536 & 0.9533 \\
  64 & 0.9642 & 0.9635 & 0.9636 & 0.9629 & 0.9627 \\
 128 & 0.9728 & 0.9713 & 0.9713 & 0.9704 & 0.9703 \\
 256 & 0.9798 & 0.9776 & 0.9776 & 0.9763 & 0.9764 \\
 512 & 0.9855 & 0.9827 & 0.9826 & 0.9812 & 0.9812 \\
1024 & 0.9901 & 0.9868 & 0.9865 & 0.9851 & 0.9850 \\
2048 & 0.9940 & 0.9899 & 0.9898 & 0.9882 & 0.9881 \\
4096 & 0.9970 & 0.9924 & 0.9927 & 0.9909 & 0.9906 \\
8192 & 0.9995 & 0.9945 & 0.9946 & 0.9923 & 0.9925 \\
\noalign{\smallskip}\hline\noalign{\smallskip}
\multicolumn{1}{l}{$\left<\varepsilon\right>$}
     & 0.03   & 0.004  & 0.0005 \\
\multicolumn{1}{l}{$\varepsilon_\mathrm{max}$}
     & 0.17   & 0.02   & 0.0026 \\
\multicolumn{1}{l}{$\left<\alpha\right>$}
     & 10.2\degree & 5.1\degree & 2.5\degree \\
\noalign{\smallskip}\hline
\end{tabular}
\begin{flushleft}
The wall potential has $c=256$ and $b=\infty$,
the atomic step size $a$ is as in Eq.~(\ref{eq:a}),
the number of sampling $n$ and burn-in $n_0$ steps is shown,
the mean $\left<\varepsilon\right>$
and largest $\varepsilon_\mathrm{max}$
local errors are from Eq.~(\ref{eq:e}),
and $\left<\alpha\right>$ is the mean bend angle
$\alpha_n = \ka_n s_{n+1}$.
The values $\varphi_\mathrm{walk}$ are from
the random walk integration,
$\varphi_\mathrm{ideal}$ are from Eq.~(\ref{eq:p0}).
\end{flushleft}
\end{table}

Table~\ref{tab:h8} shows our results
for the hot nearly-ideal few-atom gas of Hertzian spheres.
The cubic scaling of the local errors 
$\left<\varepsilon\right>$
and  $\varepsilon_\mathrm{max}$
with the step size $a$
is a clear witness of the second-order numerical accuracy
of the integrator.
We have also run the random walk integration~\cite{MRRTT53}
and the pressures $\varphi_\mathrm{walk}$ in Table~\ref{tab:h8}
match those from our isokinetic trajectory
to within the statistical errors,
which we estimate by the blocking method~\cite{FP89}.
Even though there are $3N-3$ degrees of freedom
(as the rotations of the whole are zeroed out),
there still must be the factor of $3N-1$
in the definition of $\gamma$ in Eq.~(\ref{eq:f}) ---
thanks to this few-atom example
we have learned this truth,
should we have played with many atoms from the very beginning,
we would have likely overlooked it.

\begin{table} 
\caption{Computation on $N=201$ Lennard-Jones atoms.}
\label{tab:lj}
\begin{tabular}{rcrrrrr}
\hline\noalign{\smallskip}
  \multicolumn{1}{c}{$a$}
 &
 &\multicolumn{1}{r}{$1/16$}
 &\multicolumn{1}{r}{$1/32$}
 &\multicolumn{1}{r}{$1/64$}
\\
 \multicolumn{2}{l}{$n, n_0$}
 &\multicolumn{1}{r}{$2^{25}, 2^{21}$}
 &\multicolumn{1}{r}{$2^{26}, 2^{22}$}
 &\multicolumn{1}{r}{$2^{27}, 2^{24}$}
\\
\noalign{\smallskip}\hline\noalign{\smallskip}
\multicolumn{1}{c}{$T$} & st\textsuperscript{a}
 &\multicolumn{1}{c}{$\varphi$}
 &\multicolumn{1}{c}{$\varphi$}
 &\multicolumn{1}{c}{$\varphi$}
 &\multicolumn{1}{c}{$\left<\varepsilon\right>$}
 &\multicolumn{1}{c}{$\left<\alpha\right>$}
\\
\noalign{\smallskip}\hline\noalign{\smallskip}
0.2 &  s & 0.0000 & 0.0000 & 0.0000 & 0.00054 & 9.4\degree \\
0.3 & lg & 0.0000 & 0.0020 & 0.0059 & 0.00046 & 7.5\degree \\
0.4 & lg & 0.0007 & 0.0056 & 0.0088 & 0.00044 & 6.2\degree \\
0.5 & lg & 0.0191 & 0.0307 & 0.0320 & 0.00038 & 5.2\degree \\
0.6 & lg & 0.2318 & 0.1820 & 0.1833 & 0.00032 & 4.0\degree \\
0.7 &  g & 0.8974 & 0.8975 & 0.8975 & 0.00008 & 0.6\degree \\
0.8 &  g & 0.9191 & 0.9191 & 0.9191 & 0.00007 & 0.6\degree \\
0.9 &  g & 0.9338 & 0.9337 & 0.9338 & 0.00006 & 0.5\degree \\
1.0 &  g & 0.9443 & 0.9443 & 0.9443 & 0.00006 & 0.5\degree \\
\noalign{\smallskip}\hline
\end{tabular}
\begin{flushleft}
The spherical vessel volume $V/N=64$, $c=256$, $b=1$.
Fugacity factor $\varphi$,
mean local integration error $\left<\varepsilon\right>$
and bend angle $\left<\alpha\right>$.\\
\textsuperscript{a}Phase state: (s) solid, (l) liquid, (g) gas.
\end{flushleft}
\end{table}

A more characteristic example of thermodynamic integration
is shown in Table~\ref{tab:lj} where $N=201$ Lennard-Jones atoms
in volume $V/N=64$ form solid, liquid, and gas phases
in a range of temperatures.

\section{Chemical applications}

Automated global searches for stationary points
on molecular potential energy surfaces
can be a very helpful tool for theoretical chemists,
as their chemical intuition alone can not always find
all the ways in which the atoms can bind together.
Points on the isokinetic trajectory,
at a good distance between them,
can be taken as the input to energy minimization
(or saddle point location),
and the optimized points can then be sieved
to remove duplicates and thus to find
a set of unique structures
that may be further sorted by their energies.
To get the most out of it
for a given amount of computer time,
a meaningful setting of the key parameters is needed:
the temperature $T$ should be neither too low (to overcome
the barriers sooner) nor too high (to keep the bonds unbroken),
the volume $V$ neither too small (to give some freedom of motion)
nor too big (to hold the whole together),
the distance between the points taken from the trajectory
neither too short (to get new structures most of the time)
nor too long (to keep from wasting time),
and the integration step size can now be greater but not too much.

\begin{table} 
\caption{Hit rate of global searches for stable structures
of hydrogen-bonded molecular clusters.}
\label{tab:mol}
\begin{tabular}{lrrrrrrrrrr}
\hline\noalign{\smallskip}
$T$                & \multicolumn{2}{l}{0.001}
                &&&& \multicolumn{2}{l}{0.002} \\
system             & $V/N$ & $l$ & $M$ & $m$ & $\left<k\right>$
                   & $V/N$ & $l$ & $M$ & $m$ & $\left<k\right>$ \\
\noalign{\smallskip}\hline\noalign{\smallskip}
(H$_2$O)$_5$       & 256 &    128 &  2048 &   21 & 156 & 128 &    128 &  2048 &    31 & 229 \\
(H$_2$O)$_6$       & 256 &    128 &  2048 &  143 & 170 & 128 &    128 &  2048 &   190 & 257 \\
                   & 256 &    128 & 16384 &  215 &     & 128 &    128 & 16384 &   266 &     \\
(H$_2$O)$_7$       & 256 &    128 &  2048 &  470 & 196 & 128 &    128 &  2048 &   656 & 287 \\
(H$_2$O)$_8$       & 256 &    128 &  2048 &  833 & 202 & 128 &    128 &  2048 &  1291 & 322 \\
                   & 256 &    128 & 16384 & 3006 &     & 128 &    128 & 16384 &  4640 &     \\
(H$_2$O)$_9$       & 256 &    128 &  2048 & 1150 & 220 & 128 &    128 &  2048 &  1744 & 342 \\
(H$_2$O)$_{10}$    & 256 &    128 &  2048 & 1111 & 225 & 128 &    128 &  2048 &  1920 & 371 \\
                   & 256 &    128 & 16384 & 9176 &     & 128 &    128 & 16384 & 13829 &     \\
(H$_2$O)$_{11}$    & 256 &    128 &  2048 & 1482 & 250 & 128 &    128 &  2048 &  1982 & 402 \\
(H$_2$O)$_{12}$    & 128 &     32 &  8192 & 1838 & 216 & 128 &     32 &  8192 &  5632 & 403 \\
                   & 128 &     64 &  4096 & 1564 &     & 128 &\bf  64 &  4096 &  3580 &     \\
                   & 128 &\bf 128 &  2048 & 1200 &     & 128 &\bf 128 &  2048 &  2003 &     \\
                   & 128 &\bf 256 &  1024 &  806 &     & 128 &    256 &  1024 &  1024 &     \\
                   & 256 &    128 &  2048 & 1415 & 250 & 256 &    128 &  2048 &  2007 & 543 \\
Ala(H$_2$O)$_4$    & 256 &    128 & 16384 & 1583 & 167 & 128 &    128 & 16384 &  5164 & 295 \\
Ala(H$_2$O)$_5$    & 256 &    128 & 16384 &  868 & 130 & 128 &    128 & 16384 &  9830 & 326 \\
Ala(H$_2$O)$_6$    & 256 &    128 & 16384 & 1879 & 147 & 128 &    128 & 16384 & 13107 & 353 \\
Ala(H$_2$O)$_7$    & 256 &    128 & 16384 & 3118 & 163 & 128 &    128 & 16384 & 14476 & 377 \\
Ala(H$_2$O)$_8$    & 256 &    128 & 16384 & 4419 & 174 & 128 &    128 & 16384 & 15062 & 399 \\
\noalign{\smallskip}\hline
\end{tabular}
\begin{flushleft}
Temperature $T$ and volume $V$ are in au,
$c=\frac14$, $b=1$; $a=\frac1{32}$~au. \\
Every $l$-th point from the trajectory is fed to the optimization \\
taking the mean number of steps $\left<k\right>$ to converge. \\
Among $M$ optimized points, $m$ are unique. \\
The values of $\left<k\right>$ and $m$ may vary from run to run.
\end{flushleft}
\end{table}

\begin{figure}
\includegraphics[scale=0.6]{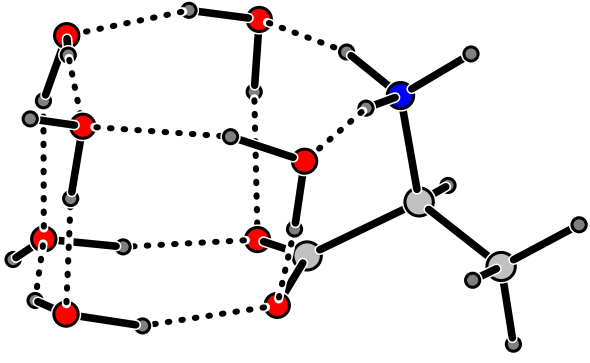}
\caption{The lowest-energy structure of alanine and six water molecules
Ala(H$_2$O)$_6$, as found by the calculations.}
\label{fig:a6w}
\end{figure}

Here we report our first experience
with such global search technique
on the example of a few hydrogen-bonded molecular clusters,
their potential energy surfaces being computed
by our parametrizable electronic structure model~\cite{L11,B17}.
First of all, we took two typical temperatures ---
0.001~au (316~K) and 0.002~au (632~K) ---
and sought, on a power-of-two scale,
the best atomic step size,
finding $\frac{1}{32}$~au to be rather good
for any molecules, with $\varepsilon_\mathrm{max} < 0.01$~au.
Next, we have tailored the other settings for the best hit rate
of the global searches, as shown in Table~\ref{tab:mol}.
On the water dodecamer (H$_2$O)$_{12}$
the distance~$l$ in steps between the to-be-optimized trajectory points
was adjusted to bring nearly the lowest cost of finding a new structure
measured as $(l + \left<k\right>)M/m$,
where $\left<k\right>$ is the mean number of steps
of a BFGS~\cite{B70,F70,G70,S70}-based energy minimization procedure,
$M$ is the number of optimized structures, $m$ of which are unique,
so $l=128$ is a good choice for both temperatures.
Doubling the volume to $V/N=256$~au
raises the cost through greater $\left<k\right>$,
more so for the higher $T$,
so we take it only in the cold case,
and with these settings we study all other molecular clusters.
A weak dependence of $\left<k\right>$ on the system size is seen.
For the smaller systems, most if not all of the stable structures
seem to be found, and the global minimum is likely to be amoung them
(for the water clusters,
the global minima reported~\cite{WH98}
for the TIP4P~\cite{BF33,JCMIK83} empirical potential
are structurally the same
as the lowest, or seldom next-to-lowest, ones found in our tests).
We cannot help showing one such lowest-energy structure
in Fig.~\ref{fig:a6w} as an illustration to the Chemist
of how this pure mathematics
opens a window to the wonderful world of wet chemistry

\section{Conclusions}

The new reversible second-order numerical integration scheme
for the isokinetic equations of motion
derived here based on the circular-arc interpolation
has an elegant simplicity of its formulas
and works well with the potential functions
typical of molecular systems.
It can also be used as a part of the hybrid~\cite{DKPR87}
stochastic-deterministic thermostats~\cite{MMT03,G03}
for systems that lack ergodicity~\cite{HB84}.

\appendix

\section{The tangent equation}
\label{sec:zeq}

To find $\vec{z}$, such that $|\vec{z}|=1$, from the equation
\begin{equation}
\label{eq:xz}
\frac{\sin(\ka s)}{\ka} \vec{z}
+ \frac{1 - \cos(\ka s)}{\ka^2}
 (1 - \vec{z}\,\vec{z} \cdot) \vec{f}
 = -\vec{x}
\end{equation}
with
\begin{equation}
\label{eq:ka}
\ka = \bigl|(1 - \vec{z}\,\vec{z} \cdot) \vec{f}\bigr|
= \sqrt{|\vec{f}|^2 - (\vec{z}\cdot\vec{f})^2}
\end{equation}
for given $\vec{f}$ and $\vec{x}$,
the dot product of both sides with $\vec{z}$ is taken first,
hence
\begin{equation}
\frac{\sin(\ka s)}{\ka} =
-\vec{z}\cdot\vec{x},
\end{equation}
from which, by the way, Eqs.~(\ref{eq:sn1}) and~(\ref{eq:sn}) follow.
Now $\sin(\ka s)$ and $\cos(\ka s)$ in terms of $\vec{z}\cdot\vec{x}$
are put back into Eq.~(\ref{eq:xz}) to get
\begin{equation}
-\vec{z}\,\vec{z}\cdot\vec{x}
+ y (1 - \vec{z}\,\vec{z} \cdot) \vec{f}
= -\vec{x}
\end{equation}
with
\begin{equation}
\label{eq:y}
y = \frac{1 - \sqrt{1 - \ka^2 (\vec{z}\cdot\vec{x})^2}}{\ka^2} ,
\end{equation}
and if rewritten as the projection
\begin{equation}
(1 - \vec{z}\,\vec{z} \cdot) (\vec{x} + y \vec{f} ) = 0,
\end{equation}
it becomes clear that
\begin{equation}
\label{eq:z}
\vec{z} = \frac{\vec{x} + y \vec{f}}{|\vec{x} + y \vec{f}|},
\end{equation}
and only $y$ is yet to be found.

To simplify further notations,
\begin{equation}
\chi = |\vec{x}|^2, \qquad
\phi = |\vec{f}|^2, \qquad
\psi = \vec{x}\cdot\vec{f},
\end{equation}
then from Eqs.~(\ref{eq:ka}) and~(\ref{eq:z}) follows
\begin{equation}
\ka^2 = \phi - \frac{(\psi + y\phi)^2}{\chi + 2y\psi + y^2 \phi}
= \frac{\chi\phi - \psi^2}{\chi + 2y\psi + y^2 \phi},
\end{equation}
putting it together with
\begin{equation}
(\vec{z}\cdot\vec{x})^2 = \frac{(\chi + y\psi)^2}{\chi + 2y\psi + y^2 \phi}
\end{equation}
into Eq.~(\ref{eq:y}) yields
\begin{equation}
\sqrt{1 - \frac{(\chi\phi - \psi^2)(\chi + y\psi)^2}{(\chi + 2y\psi + y^2 \phi)^2}}
= 1 - \frac{(\chi\phi - \psi^2) y}{\chi + 2y\psi + y^2 \phi},
\end{equation}
multiplying both sides with $\chi + 2y\psi + y^2 \phi$
followed by squaring, gathering the terms,
and dividing by $\chi\phi - \psi^2$ leads to
\begin{equation}
2(\chi + 2y\psi + y^2 \phi) y - (\chi\phi - \psi^2) y^2 - (\chi + y\psi)^2
= 0
\end{equation}
and at last to the cubic equation
\begin{equation}
2\phi y^3 + (4\psi - \chi\phi) y^2 + 2\chi(1 - \psi) y - \chi^2 = 0,
\end{equation}
looking at which we see that it always
(for any $\phi$ and $\psi$) has the root $y=\frct12\chi$,
and this is the only real root,
except the unlikely special case $\psi^2 = \chi\phi$ of no interest,
as can be shown by rewriting the cubic equation as
\begin{equation}
(2 y - \chi) (\phi y^2 + 2\psi y + \chi) = 0
\end{equation}
and knowing the inequality $\psi^2 \le \chi\phi$,
thus the answer is
\begin{equation}
\vec{z} =
\frac{\vec{x} + \frct12 |\vec{x}|^2 \vec{f}}
{\left|\vec{x} + \frct12 |\vec{x}|^2 \vec{f}\right|} .
\end{equation}

Finding $\vec{f}$ for given $\vec{x}$ and $\vec{z}$ leads to
\begin{equation}
\vec{f} = 2\frac{u\vec{z} - \vec{x}}{|\vec{x}|^2}
\end{equation}
with arbitrary $u$, but if this $\vec{f}$ is then needed
in projected form $(1 - \vec{z}\,\vec{z} \cdot) \vec{f}$
as in Eq.~(\ref{eq:c}), then $u = 0$ is a natural choice,
and
\begin{equation}
\label{eq:fx}
\vec{f} = -2\frac{\vec{x}}{|\vec{x}|^2} .
\end{equation}

\section{The step length equation}
\label{sec:seq}

To solve the equation
\begin{equation}
u \sin^2 \theta + 2 w (1 - \cos \theta) \sin \theta
+ v (1 - \cos \theta)^2 = 1,
\end{equation}
a change to variable
\begin{equation}
s = \sin^2 \left(\frct12 \theta \right)
\end{equation}
makes
\begin{equation}
4 u s (1 - s) + 8 w s \sqrt{s - s^2} + 4 v s^2 = 1,
\end{equation}
moving the terms to leave the square root alone
on the left hand side, then taking the square of both sides,
\begin{equation}
64 w^2 (s^3 - s^4) = \left( 1 - 4 u s + 4 (u - v) s^2 \right)^2 ,
\end{equation}
and gathering the powers of $s$
yields the quartic equation 
\begin{eqnarray} \nonumber
1 &-& 8us + 8 \left(u - v + 2u^2 \right) s^2
 - 32 \left(u (u - v) + 2w^2 \right) s^3 \\
&+& 16 \left((u - v)^2 + 4w^2 \right) s^4 = 0.
\end{eqnarray}
Its straightforward solution by standard means
is not advisable, however,
as it can easily suffer from rounding errors and
floating-point overflow in machine computation.

Another change of variable
\begin{equation}
x = \frac1{u + v}\left(\frac1s - 2 u\right)
= \frac1{u + v}
\left(\frac1{\sin^2 \left(\frct12 \theta \right)} - 2 u\right)
\end{equation}
leads to the well-behaved depressed quartic equation
\begin{eqnarray} \nonumber
x^4 &+& 8\frac{u - v - u^2}{(u + v)^2} x^2
- 64\frac{w^2}{(u + v)^3} x \\
&+& 16\frac{(u - v - u^2)^2 + (4 - 8u)w^2}{(u + v)^4} = 0
\end{eqnarray}
that can be stably solved by the classic 16th--18th century methods
on modern hardware.

The quartic equation
\begin{equation}
u s^2 + w s^3 + v s^4 = 1
\end{equation}
that comes from Eq.~(\ref{eq:qin})
needs careful handling when $w\ll u^{3/2}$ and $v\ll u^2$,
which often happens.
With a good threshold value $\eta\approx 2^{-10}$,
if
\begin{equation}
w > \eta u^{3/2}
\quad\mathrm{or}\quad
v > \eta u^2,
\end{equation}
then a change of variable
\begin{equation}
s = h/x,
\end{equation}
\begin{equation}
h = \frac1{\sqrt[4]{u^2 + v}},
\end{equation}
yields a well-behaved depressed quartic equation
\begin{equation}
x^4 - h^2 u x^2 - h^3 w x - h^4 v = 0
\end{equation}
that can be solved by standard means
with some care taken to gather the terms
in the determinant of its resolvent cubic equation.
Otherwise, the change of variable
\begin{equation}
s = \frac1{\xi\sqrt{u}}
\end{equation}
leads to
\begin{equation}
\xi^4 - \xi^2 - \gamma \xi - \beta = 0,
\end{equation}
with
\begin{equation}
\gamma = \frac{w}{u^{3/2}}, \qquad
\beta = \frac{v}{u^2},
\end{equation}
and now the function $\xi(\gamma,\beta)$
is expaned in powers of its arguments
by perturbation theory,
\begin{equation}
\xi = 1 + \frct12 \gamma + \frct12 \beta
- \frct38 \gamma^2 - \gamma\beta - \frct58 \beta^2
+ \frct12 \gamma^3 + \dots,
\end{equation}
the terms up to $\gamma^3$, as written,
are enough to get a good accuracy.

Last but not least, the identity
$1-\cos\theta = 2\sin^2\left(\frct12 \theta\right)$
should be used for a stable evaluation of $1-\cos\theta$,
in Eq.~(\ref{eq:c}) and everywhere, in numerical computations.

\section{Inclusion of masses and comparison
to the velocity-rescaling scheme}
\label{sec:mass}

For $N$ atoms with masses $\{\mm_i\}$ and positions $\{\vec{\mr}_i(\mt)\}$
in time $\mt$, the isokinetic equations of motion
\begin{equation}
\label{eq:mkin}
\mm_i \frac{\de^2 \vec{\mr}_i(\mt)}{\de\mt^2} =
\vec{\mf}_i \bigl(\vec{\mr}(\mt)\bigr)
- \eta(\mt) \frac{\de \vec{\mr}_i(\mt)}{\de\mt},
\end{equation}
\begin{equation}
\label{eq:mf}
\vec{\mf}(\vec{\mr}) =
-\frac{\partial U(\vec{\mr})}{\partial\vec{\mr}},
\end{equation}
look like the Newton equation
with the time-dependent friction coefficient
(that can be negative)
\begin{equation}
\eta(\mt) = \frac1{2K(\mt)}\,
 \frac{\de \vec{\mr}(\mt)}{\de\mt}
 \cdot \vec{\mf} \bigl(\vec{\mr}(\mt)\bigr)
\end{equation}
which, by construction, conserves the kinetic energy
\begin{equation}
K(\mt) = \frct12 \sum\limits_i^N \mm_i
 \left| \frac{\de \vec{\mr}_i(\mt)}{\de\mt} \right|^2
\end{equation}
that can be connected to the temperature
\begin{equation}
\label{eq:mkt}
K(\mt) = \frct12 (3N-1)T .
\end{equation}
Using the mass-weighted coordinates
\begin{equation}
\vec{r}_i = \sqrt{\mm_i} \;\vec{\mr}_i
\end{equation}
and the temperature-scaled time
\begin{equation}
t = \sqrt{(3N-1)T}\;\mt
\end{equation}
Eqs.~(\ref{eq:mkin}) and~(\ref{eq:mf})
become Eqs.~(\ref{eq:kin}) and~(\ref{eq:f}),
and Eq.~(\ref{eq:mkt}) is now simply $|\dot{\vec{r}}(t)| = 1$.

Thus the easiest and most efficient implementation
of our integrator into a computer code that works with masses
would be (a) the conversion to the scaled coordinates,
(b) the work of Section~\ref{seq:theory},
and (c) the back-conversion to the physical coordinates.

Careful reading of Ref.~\cite{W72} lets us write
its working equations as
\begin{eqnarray}
\nonumber
\vec{\mr}_i (\mt + \Delta\mt) &=& \vec{\mr}_i (\mt)
+ \bigl(\vec{\mr}_i (\mt) - \vec{\mr}_i (\mt - \Delta\mt)\bigr) \sigma (\mt)
\\
&+& \frac{(\Delta\mt)^2}{\mm_i} \vec{\mf}_i \bigl(\vec{\mr} (\mt)\bigr)
\end{eqnarray}
with the velocity-rescaling factor
\begin{equation}
\sigma^2 (\mt) =
\frac{(3N-1)T(\Delta\mt)^2}{\sum\limits_j \mm_j \bigl|\vec{\mr}_j (\mt) - \vec{\mr}_j (\mt - \Delta\mt) \bigr|^2}
\end{equation}
simplified to be global for all mass species
and with the number of degrees of freedom adjusted from $3N$ to $3N-1$.
In the scaled variables it becomes
\begin{equation}
\vec{r} (t + \Delta t) = \vec{r} (t)
+ \Delta t
 \frac{\vec{r} (t) - \vec{r} (t - \Delta t)}
      {\bigl|\vec{r} (t) - \vec{r} (t - \Delta t) \bigr|}
+ (\Delta t)^2 \vec{f} \bigl(\vec{r} (t)\bigr),
\end{equation}
its time-reversal asymmetry is clearly seen.
In the notations of Section~\ref{seq:theory}
it can be written as
\begin{equation}
\label{eq:vr}
\vec{x}_{n+1} =
 \tau \frac{\vec{x}_n} {|\vec{x}_n|}
+ \tau^2 \vec{f}_n,
\end{equation}
with $\tau\equiv \Delta t$,
and compared to our Eqs.~(\ref{eq:xrf}) and~(\ref{eq:zn}).
In both cases, the next step $\vec{x}_{n+1}$
lies in the space spanned by vectors
$\vec{x}_n$ and $\vec{f}_n$,
but whereas our integrator moves on the circle
with unit ``velocity'' and thus conserves
the fundamental isokinetic property of the trajectory
also between the discretization points,
the velocity-rescaling~\cite{W72} integrator,
as seen in Eq.~(\ref{eq:vr}),
moves with ``acceleration'' on the parabola
and runs faster away from the exact solution
as the step size grows.

\bibliographystyle{spphys}
\bibliography{isokin_arxiv2}

\clearpage

\textbf{Timeline of Comments from Journal Editors and Reviewers}\\

\noindent
1.~\textbf{The Journal of Chemical Physics}
\begin{tabbing}
Received: \= 17-Sep-2017 \\
Rejected: \> 07-Dec-2017 
\end{tabbing}
\textit{Editor's message}:
\begin{verbatim}
In light of the recommendations of the reviewers, we have decided
not to accept your manuscript in <I>The Journal of Chemical Physics</I>
because it does not fulfill at least one of the core criteria for acceptance.
 
However, with sufficient revisions, your manuscript may be suitable
for consideration in <I>AIP Advances</I>, another AIP Publishing journal.
<I>AIP Advances</I> is an open access journal focused on rapid publication
of original, scientifically sound articles in all areas of the physical
sciences. For more information about the Journal visit
http://aip.scitation.org/adv/info/focus. 
 
If you wish to submit your manuscript to <I>AIP Advances</I>,
simply click on the link below and follow the instructions.
Once the transfer process is complete you will receive an email
inviting you to upload a revised version of your manuscript
to <I>AIP Advances</I>. You should make every attempt to address
the comments of the reviewers and upload the revised version
of your manuscript together with a point-by-point response
to the reviewer comments. The reviewer reports will be transferred
with your manuscript, enabling the <I>AIP Advances</I> Editors
to quickly review your revised manuscript to reach an <B>informed,
rapid decision.</B> 
 
For your convenience, please find the reviewers' reports included below.
 
Thank you for the opportunity to examine this work.
\end{verbatim}

\noindent
\textit{Reviewers' comments}:
\begin{verbatim}
Reviewer #1 Evaluations: 
Recommendation: Reject 
New Potential Energy Surface: No 
 
Reviewer #1 (Comments to the Author): 
 
The manuscript by Laikov reports a new, reversible MD integrator
for isokinetic simulations. 
 
I, unfortunately, do not recommend publication at this time. 
 
1. The writing is, frankly, rather strange. The style is not suitable
for JCP, and the depth of analysis is not sufficient to convince
a reader of the applicability of the algorithm. 
 
2. The new integrator may, in fact, be correct, and I do not see
any formal objection to the method. I have not tested its implementation
myself. However, the analysis, at the very least, contains several
typographical errors that make assessment challenging. It also ignores
past integrators of Tuckerman/Martyna and Ely (and likely others). 
 
3. The lone connection to chemical physics is the search for isomers
of hydrated alanine, for which almost no details were provided.
(I was somewhat surprised that the zwitterionic form is more stable
with only six water molecules present. A quick set of computations
showed that this conclusion is likely true, although it is also
highly sensitive to the level of theory.) More significant is the fact
that this demonstration proved essentially nothing in the present manuscript.
Was this isomer historically difficult to find? Did other isomer-search
protocols (Monte Carlo, genetic algorithms, etc.) somehow fail here?
Is the present protocol more efficient than these options? 
 
 
Reviewer #2 Evaluations: 
Recommendation: Major revision 
New Potential Energy Surface: No 
 
Reviewer #2 (Comments to the Author): 
 
Report on the manuscript A17.09.0159
"A reversible numerical integrator of the isokinetic equations of motion",
by Dimitri N. Laikov. 
 
This manuscript addresses much more than a purely numerical issue
although it might look at first like a mere numerical piece of research.
Newtonian mechanics and many-particle ensembles and interactions
are some of the key points the manuscript tackles with the purpose being
to advance, by means of a reversible integrator, the cause of numerical
wet chemistry. The subject-matter is modern and important for chemists
and physicists. It is even more so for those working at the frontier
of these two areas of research and, more particularly, in soft-matter science
since the method being developed by the author should be adaptable
to a wide palette of situations encountered in out-of-equilibrium
Molecular Dynamics simulations. In fact, non-equilibrium Molecular Dynamics,
NEMD is a fascinating research topic. The purpose of the scientists
working in it is the study of many-particle ensembles whose evolution
is driven away from equilibrium by the mere presence of the action
of an external field. It is precisely the absence of equilibrium
which makes the role of thermostats so important for this kind of studies.
Mathematically, the isokinetic equations of motion are a pivotal concept
which becomes the focus of such simulations in order for the kinetic energy
of the many-particle ensembles to be dynamically constrained during the run.
It is also so in this manuscript. How to use circular arcs passing
through each 3 consecutive points along a trajectory is, synoptically,
what the present manuscript tells about. According to the author's sayings,
the method outlined in this manuscript is especially robust and allows
a good control of the variable step length. 
 
After reading and rereading this manuscript, the overall impression
I have left with is both a satisfaction by the high level of research
which I feel this is all about, and at the same time a feeling
of unfinished work. To be honest, I had a rather hard time&#x00A0;reading
this manuscript and I spent much time to digest its content.
Much of the difficulty lies with language. In fact, the awkward use of English
in a research paper may quickly confuse or tire the reader,
and more importantly obscure the core message. It is probably the case here.
I had a serious difficulty in comprehending what is being said by the author
in several places throughout the manuscript (many sentences are too lengthy
to still have a meaning; see for instance Introduction section, but also
in Abstract and elsewhere). The understanding of the paper as a whole
is problematic-I would say, in its present form. The reduced explanations
offered by the author throughout his presentation did not help me to do better
or to counter this difficulty. This applies to practically all Sections
(Introduction and Theory, Chemical Applications, ...).
Although short manuscripts are generally welcomed for consideration,
in the present case the manuscript should be somewhat expanded, in my opinion,
and profitably be (re)organized in order for the focus to be put where
it needs to be. Introduction is too short a section in the present form.
The author should pay all due care to properly develop this section
(~1 page or so) in order to better guide the reader to fully understand
the essence of the problem treated throughout the manuscript and the reason
to why the author's method is superior to what previous algorithms are also
supposed to do. Is it possible that the author makes (numerical) comparisons
with available integrators? If yes, this would be particularly welcomed
as more than any other discussions it should best highlight the advantage
of the method being developed. Bibliography is very limited and should be
enriched and expanded. There are several flagship references presently missing.
Here is few of them: Statistical Mechanics of nonequilibrium solids
(chapter on linear response theory, equilibrium distribution functions, etc.),
Minary et al jcp 118, 2510 (2003),(...) Eq. (1) should be reformulated.
The presence of a hardly-visible dot product at the right-hand side
of the fraction can only create confusion. There exist better notations
in literature. Please, set masses rather than considering m=1
(or use momenta instead). The author states "it seems natural to take
the weights ..." (equation 15). Where does this "natural assessment" come from?
Is there evidence for that? The author should explain in detail the purpose
and the advantages of elaborating circular curve interpolation (equations 4-7).
Likewise, special emphasis needs to be put on the reversibility property
of the computed trajectories. Although reversibility plays a crucial role
in the new numerical integrator (the concept appears even in the Title!
of the manuscript), I see almost no mention of that concept (with the exception
of one italicized occurrence). The author should clearly distinguish between
(or, instead (where relevant), avoid to use) different notations for the same
quantity. For instance, r_n(t) and r(t_n),... A cartoon illustrating
some vectors such as x_n, z_n along with the interpolated arcs would be
particularly welcomed in order for the reader to immediately grasp the point.
TABLES II and III are too technical a material for my taste. I would prefer
a more physical property which would clearly indicate the significance
of the study. 
Exclamatory sentences have no place in a scientific text intended
for publication even though I can understand that by using words
such as wow! etc, the author wishes to&#x00A0;share his surprise&#x00A0;and
concern throughout his document. Nor should use&#x00A0;be
made&#x00A0;of&#x00A0;superlatives&#x00A0;or of everyday
language&#x00A0;in such documents. 
 
In conclusion, I think the author has&#x00A0;some interesting material
to share. However, the manuscript by no means is suitable for the jcp
in its present form nor could it be so after a minor revision.
Substantial reconsideration is necessary before publication could be envisaged.
 
Recommendation: Major revision and second round of reviewing. 
\end{verbatim}

\vspace{1em}
\noindent
2.~\textbf{Journal of Chemical Theory and Computation}
\begin{tabbing}
Received: \= 10-Dec-2017 \\
Rejected: \> 02-Jan-2018 
\end{tabbing}
\textit{Editor's message}:
\begin{verbatim}
Thank you for considering the Journal of Chemical Theory and Computation
for your manuscript submission. The Journal receives far more papers
than it can publish.  After initial screening and editorial review,
it has been decided to return your paper without further review.
The submission is interesting, but there is no demonstration that
the new method represents an advance over existing alternatives.
The  manuscript is better suited for a journal that is more central
to mathematical chemistry or fundamental liquid-state theory.
I regret any inconvenience, but I hope that  you will consider
alternative journals without further delay.
\end{verbatim}

\vspace{1em}
\noindent
3.~\textbf{Journal of Computational Chemistry}
\begin{tabbing}
Received: \= 04-Jan-2018 \\
Rejected: \> 04-Jan-2018 
\end{tabbing}
\textit{Editor's message}:
\begin{verbatim}
I am sorry to inform you that your manuscript, A reversible numerical
integrator of the isokinetic equations of motion, has not been found acceptable
for publication in Journal of Computational Chemistry.  JCC publishes
only papers which report about progress in the Computational Chemistry part
that is interesting for the broad community of the readers. Research
which focuses on a very specialized field using standard methods of theory
should better become published in a more specialized journal. I suggest
that you submit your work to a journal in the field of mathematical chemistry.
\end{verbatim}

\vspace{1em}
\noindent
4.~\textbf{Molecular Physics}
\begin{tabbing}
Received: \= 05-Jan-2018 \\
Rejected: \> 22-Feb-2018 \\
Revised:  \> 26-Feb-2018 \\
Rejected: \> 19-Mar-2018 
\end{tabbing}
\textit{Editor's message}:
\begin{verbatim}
Your manuscript "A reversible numerical integrator of the isokinetic equations
of motion" which you submitted to Molecular Physics has now been reviewed
and the referees comments are included at the bottom of this email. 

You will see that both referees recommend substantial revision;
we hope that you will be able to undertake this and look forward to receiving
a revised manuscript in due course.

When you revise your manuscript, please highlight the changes you make
in the manuscript by using the track changes mode in MS Word,
or by using bold or coloured text.

To start the revision, please click on the link below:
\end{verbatim}
\textit{\tiny(technical details follow)}
\begin{verbatim}
Thank you again for submitting your manuscript to Molecular Physics
and we look forward to receiving your revision.
\end{verbatim}
\textit{Reviewers' comments}:
\begin{verbatim}
Referee: 1

Comments to the Author

The authors introduce a new reversible integrator for the isokinetic equations
of motion and apply it to several test cases.  The work might be publishable,
but major revisions will be needed to address a number of key questions.

The first question is what is the motivation for proposing this new integration
scheme?  For example, how does it compare to the integrator in Ref. 11?
Does it offer an advantage over that one, or is it just an equivalent
alternative?  If the instantaneous kinetic energy is plotted as a function
of time for this integrator, how well is it conserved?  Note that the isokinetic
equations of motion exactly conserve the kinetic energy by construction,
and the integrator should as well.

Second, does this new integrator converge observable properties faster
than a standard thermostat, such as Langevin or Nose'-Hoover chains would?
Such a convergence comparison is needed.

Third, the authors only show global thermodynamic properties for the particles
in the spherical vessel.  Can the integrator reproduce accurate spatial
distributions?  The authors should generate a radial distribution function
for these particles and compare the result to a benchmark one generated
using the same standard thermostat suggested above.  

Finally, on a minor point, I would suggest that the authors show
how their integrator is formulated in unscaled gradients,
as these are what most molecular dynamics codes use.



Referee: 2

Comments to the Author

Manuscript Reference:  TMPH-2018-0012

Manuscript Title: A reversible numerical integrator of the isokinetic
equations of motion

Manuscript Author: Laikov, Dimitri

There is no doubt about the importance of undertaking molecular dynamics
simulations at constant temperature. The author develops an approach
based on the isokinetic equations of motion to facilitate simulations
at constant temperature.

Though the methodology could be of scientific interest to readers
of Molecular Physics and the scientific community in general,
the current paper falls short in a major respect: 
the lack of comparisons with other isothermal algorithms. My main concern
is that there is no way of assessing the validity of the method
without providing detailed comparisons with, for example the standard
Nose-Hoover algorithm or even simple velocity rescaling.

I am afraid that I cannot recommend consideration of the manuscript
for publication in Molecular Physics unless the author provides
this type of comparison.
\end{verbatim}
\textit{Author's reply}:
\begin{verbatim}
Dear Editor and Referees,

Many thanks for your kind attention and the suggestions
for improving the quality of my manuscript.
I have revised it to address the most important points
raised in the comments.
The changes and additions are marked in blue color.
After some thinking, I have decided to write Appendix C
(and a sentence in the Theory section that refers to it),
where the equations with masses are treated
and also a comparison with the original
velocity-rescaling integrator is made.
I have also added a few words in the Theory section
to note the ergodicity problem,
and also in the Conclusions, with two more references
added, one is about the ergodicity problem [23]
and the other is the classic work
on the Hybrid Monte Carlo [22].

I have also found that the classical 1984 work
of Suichi Nosé was published both in the J. Chem. Phys.
http://dx.doi.org/10.1063/1.447334
and in the Mol. Phys.
http://dx.doi.org/10.1080/00268978400101201
and that the Mol. Phys. publication is earlier,
so I have replaced Ref. 8 in favour of the earlier report.

Below are my answers to the Referee's Comments
(quoted with ">").

Sincerely yours,
Dimitri Laikov


> Referee's Comments to Author:
>
> Referee: 1
>
> Comments to the Author
>
> The authors introduce a new reversible integrator
> for the isokinetic equations of motion and apply it
> to several test cases.
> The work might be publishable, but major revisions
> will be needed to address a number of key questions.
>
> The first question is what is the motivation
> for proposing this new integration scheme?

From a purely mathematical point of view,
the discovery of a new numerical method
for solving a known kind of differential equations
is already enough motivation for doing it
and also enough for publishing it.

> For example, how does it compare to the integrator in Ref. 11?
> Does it offer an advantage over that one,
> or is it just an equivalent alternative?

In Ref. 11
http://dx.doi.org/10.1063/1.1534582
entitled
"Algorithms and novel applications based on the isokinetic ensemble.
I. Biophysical and path integral molecular dynamics"
the numerical integrator invented by Dr. Zhang
cited by me as Ref. 10
http://dx.doi.org/10.1063/1.473273
is thoroughly tested on wide classes of systems.
And this is the point!
Dr. Zhang did limited tests of the integrator
in the original publication, and then the other researches
did extended tests and found novel applications.
I hope that my limited tests are enough to see
that my new integrator works, but how well it compares
with the other ones might be the subject of further studies,
which are very welcome.

> If the instantaneous kinetic energy is plotted
> as a function of time for this integrator,
> how well is it conserved?

It is conserved exactly, by construction,
and not only at the discretisation points,
but also in between (on the circular interpolation curves).
The latter is not the case for the other known integrators.
This is noted once again in the text I added in Appendix C.

> Note that the isokinetic equations of motion
> exactly conserve the kinetic energy by construction,
> and the integrator should as well.
>
> Second, does this new integrator converge observable
> properties faster than a standard thermostat,
> such as Langevin or Nose'-Hoover chains would?
> Such a convergence comparison is needed.

My integrator is new, but it integrates the old (known)
differential equations.

This question is about the convergence
of the isokinetic trajectory itself, and an answer to it
should not depend on the specific numerical integrator
as soon as it works and the step size is small enough
to get an accurate trajectory.
It is known that for some systems the isokinetic trajectory
is not ergodic... so it does not converge
to the canonical distribution at all.
I added a note about this and also the Ref. 23
that deals with the question of non-ergodicity.

> Third, the authors only show global thermodynamic
> properties for the particles in the spherical vessel.
> Can the integrator reproduce accurate spatial distributions?
> The authors should generate a radial distribution function
> for these particles and compare the result to a benchmark one
> generated using the same standard thermostat suggested above.

I think that such extended tests would make my article
too long and lead too far away from its main topic,
but it could be the subject of a further study
that can be done and published by others.

> Finally, on a minor point, I would suggest that the authors
> show how their integrator is formulated in unscaled gradients,
> as these are what most molecular dynamics codes use.

The Appendix C is now added that shows how to do it,
and a sentence at the end of the Theory section invites
to read it.

> Referee: 2
>
> Comments to the Author
>
> Manuscript Reference:  TMPH-2018-0012
>
> Manuscript Title: A reversible numerical integrator of the isokinetic equations of motion
>
> Manuscript Author: Laikov, Dimitri
>
> There is no doubt about the importance of undertaking molecular dynamics simulations 
> at constant temperature. The author develops an approach based on the isokinetic equations 
> of motion to facilitate simulations at constant temperature.
>
> Though the methodology could be of scientific interest to readers of Molecular Physics and
> the scientific community in general, the current paper falls short in a major respect: 
> the lack of comparisons with other isothermal algorithms.

If the comparison in numerical tests on some model system is meant,
then I do make such comparison with the Monte-Carlo algorithm in Table 1.
If the comparison of mathematical equations is meant,
then this is addressed in the now added Appendix C.

> My main concern is that there is no way of assessing the validity of the method
> without providing detailed comparisons with, for example the standard
> Nose-Hoover algorithm or even simple velocity rescaling.
>
> I am afraid that I cannot recommend consideration of the manuscript for publication in
> Molecular Physics unless the author provides this type of comparison.

I do believe in the internal consistency of mathematics.

There is the pure mathematical object -- the isokinetic equations of motion,
a differential equation.
I find another pure mathematical object -- a numerical integrator
for this differential equation.
To "assessing the validity of the method" one can use
the tools of mathematical analysis alone,
such as limits, Taylor series, and so on.
One can also run numerical tests and study the behaviour
of the integrator for smaller and bigger step sizes.
(I do the latter in my tests both on simple model pair potentials
as well as on some quantum-chemical potentials of molecular systems.)

If we assume "that there is no way of assessing the validity of the method
without providing detailed comparisons with... the standard algorithm",
then this argument could have been applied to all methods discovered before,
also to the very first of them, but this first one
could not be compared to anything at the time of its discovery,
so its "validity" could not be assessed, and so all the following methods
could only be compared to something without the assessed validity.
Is it a paradox or a contradiction?
Or is the word "standard" a key?
Why is the "simple velocity rescaling" a standard?

I can agree that the comparison is always good.
So I added Appendix C, with the comparison
to the velocity-rescaling, in the analytical form.
\end{verbatim}
\textit{Editor's message}:
\begin{verbatim}
Thank you for submitting your response to the referees and revised manuscript,
"A reversible numerical integrator of the isokinetic equations of motion"
which you submitted to Molecular Physics.

Both referees feel that you have not made the effort to address the important
points that they raised. Could I please ask you to make a proper attempt
to respond to all of the issues in the previous referee reports (appended again
below) and revise the manuscript accordingly.

If you are not able to follow the suggestions made by the referees then
I am afraid that we will not be able to reconsider your paper for publication.

When you revise your manuscript, please highlight the changes you make
in the manuscript by using the track changes mode in MS Word or by using
bold or coloured text.
\end{verbatim}
\textit{\tiny(technical details follow)}
\begin{verbatim}
Thank you again for submitting your manuscript to Molecular Physics,
and we look forward to receiving your revision.
\end{verbatim}
\textit{Reviewers' comments}: \textit{(SILENCE)}

\vspace{1em}
\noindent
5.~\textbf{Physical Chemistry Chemical Physics}
\begin{tabbing}
Received: \= 23-Mar-2018 \\
Rejected: \> 27-Mar-2018 
\end{tabbing}
\textit{Editor's message}:
\begin{verbatim}
Thank you for your recent submission to Physical Chemistry Chemical Physics,
published by the Royal Society of Chemistry. All manuscripts are initially
assessed by the editors to ensure they meet the criteria for publication
in the journal.

After careful evaluation of your manuscript, I regret to inform you that
I do not find your manuscript suitable for publication in Physical Chemistry
Chemical Physics as it is not of sufficiently broad appeal for the general
physical chemistry readership of the journal. Therefore your article
has been rejected from Physical Chemistry Chemical Physics.

Full details of the initial assessment process can be found at:
http://www.rsc.org/Publishing/Journals/guidelines/AuthorGuidelines/JournalPolicy/initialassessment/index.asp

I am sorry not to have better news for you, however, thank you for giving
Physical Chemistry Chemical Physics the opportunity to consider your manuscript.
I wish you every success in publishing this manuscript elsewhere.
\end{verbatim}

\vspace{1em}
\noindent
6.~\textbf{Journal of Computational Physics}
\begin{tabbing}
Received: \= 29-Mar-2018 \\
Rejected: \> 07-Apr-2018 
\end{tabbing}
\textit{Editor's message}:
\begin{verbatim}
Thank you for your submission. I have carefully examined it, but I regret
to inform you that I do not believe that it is suitable for the Journal
of Computational Physics. The Journal generally publishes articles on new
or improved numerical methods, with examples showing the utility
of the proposed approach. We have not published much in the subject area
of your manuscript recently and the manuscript is therefore unlikely
to reach its target audience in the Journal. Instead of going through
the full review cycle, which is likely to take several months and probably
results in the rejection of your manuscript, I have decided to return it
to you so you can submit it to a more suitable journal. Please note
that my decision in no way reflects a judgment on my part about the quality
of your manuscript, only its suitability for the Journal of Computational
Physics.

I am sorry I cannot accept your manuscript in the Journal of Computational
Physics, and I hope your work will be published in an appropriate journal
where it will reach the audience it deserves.

Please note that we do not reconsider manuscripts that have previously been
rejected.
\end{verbatim}

\vspace{1em}
\noindent
7.~\textbf{Chemical Physics Letters}
\begin{tabbing}
Received: \= 10-Apr-2018 \\
Rejected: \> 12-Apr-2018 
\end{tabbing}
\textit{Editor's message}:
\begin{verbatim}
Thank you for submitting your manuscript to Chemical Physics Letters.
Your paper, referenced above, has now been reviewed, and we regret
to inform you that publication in Chemical Physics Letters
has not been recommended.

The comments we have received on your paper are attached below,
and we hope that these will be useful.

Thank you again for having thought of Chemical Physics Letters.

However, I do think it could be considered by another journal,
and I would like to suggest that you take advantage of the article
transfer service that "Chemical Physics Letters" is part of.
This gives you the option to have your manuscript files and details
transferred to another journal. This removes the need for you
to resubmit and reformat your manuscript, saving you valuable time
and effort during the submission process.

If you click the link below you will find relevant information
about the journal(s) to which I recommend transferring your submission.
You have the option to accept or decline the transfer offer
from the same web page:
https://ees.elsevier.com/cplett/l.asp?i=

This offer does not constitute a guarantee that your paper will be published
in the suggested Journal, but it is our hope that this arrangement will help
expedite the process for promising papers.

To learn more about the new article transfer service, please visit
www.elsevier.com/authors/article-transfer-service
\end{verbatim}
\textit{Reviewer's comments}:
\begin{verbatim}
Reviewer comments:
This manuscript is more suitable for a professional journal than CPL.
\end{verbatim}

\vspace{1em}
\noindent
8.~\textbf{Chemical Physics}
\begin{tabbing}
Received: \= 16-Apr-2018 \\
Rejected: \> 20-Apr-2018 
\end{tabbing}
\textit{Editor's message}:
\begin{verbatim}
Thank you for your submission to Chemical Physics. Having now evaluated
your manuscript, I feel there are other Elsevier journals that may better suit
the scope.
Please click on the link below to find out more about the alternative journals
I recommend. They all participate in Elsevier’s Article Transfer Service,
which means you do not have to resubmit or reformat your manuscript,
you just have to accept or decline the transfer offers. The link is valid
for 30 days.
\end{verbatim}
\textit{The list of recommended journals included\dots}
Chemical Physics Letters (!)
\textit{and I have chosen to transfer my article there.
It did not show up there again\dots
The ring was closed.}

\vspace{1em}
\noindent
9.~\textbf{International Journal of Quantum Chemistry}
\begin{tabbing}
Received: \= 25-Apr-2018 \\
Rejected: \> 11-May-2018 
\end{tabbing}
\textit{Editor's message}:
\begin{verbatim}
Thank you for submitting your manuscript, "A reversible numerical integrator
of the isokinetic equations of motion", to the Int. Journal of Quantum Chemistry.

Having considered the work in detail, I'm sorry to say that we are unable
to offer to publish it in the journal. Since we receive many more manuscripts
than we can possibly publish, it is our practice to return a proportion
without external reviews. Such decisions are made by the editors to minimize
delays for authors in cases where the paper is considered unlikely to be
appropriate for publication in the journal, even if verified as technically
correct by the referees.

In this case, we don't doubt that your study of mathematical chemistry
will be of direct interest to other specialists. However, we are sadly unable
to conclude that the work is likely to be of immediate interest to our broader
audience from all parts of the quantum chemistry community. We therefore feel
that the manuscript would find a more appropriate outlet in a more specialized
journal.

I'm sorry we can't be more positive on this occasion, but thank you
for submitting your manuscript, and I hope the outcome of this specific
submission will not discourage you from sending future manuscripts to us.

I wish you every success in rapid publication of the work elsewhere.
\end{verbatim}

\vspace{1em}
\noindent
10.~\textbf{Physical Review E}
\begin{tabbing}
Received: \= 20-May-2018 \\
Rejected: \> 06-Sep-2018 
\end{tabbing}
\textit{Unusual message}:
\begin{verbatim}
Re: Review_request HOOVER ES11735 Laikov
From: william hoover <hooverwilliam@yahoo.com>
  To: pre@aps.org
  CC: laikov@rad.chem.msu.ru
  Date: Wed Jun 20 21:33:09 2018
  Attachments: LAIKOV.pdf LAIKOV2.pdf
   
Dear Alexander Wagner,

I think that the paper has merit and interest as time-reversibility
is fundamental to simulation and, so far as is known, cannot be achieved
precisely in nonequilibrium simulations, which are possible using isokinetic
thermostatting equations of motion.  The question which needs to be clarified
is the order of the error which can be achieved with the proposed algorithm.
  The author should compare his approach to the approach indicated
in the first equation on page 4553 of LAIKOV2 ( attached above ).  In order
to make such a comparison a problem which allows an isokinetic solution
but which is also as simple as possible is required.  I suggest that the author
solve the cell-model problem described in LAIKOV.pdf ( attached above )
and use the criteria in that paper to describe the quality of his time
reversibility.  For a truly time-reversible algorithm integer arithmetic
is necessary ( but not sufficient ) and this is something on which the author
might wish to comment or demonstrate.

In summary I recommend that the author include his solution to the cell-model
problem of LAIKOV.  I would be happy to review the paper with this addition
and believe that it would be worth publishing in Physical Review E in view
of the interest of time-reversibility and thermostatted equations of motion
apparent in the references.  I am taking the liberty of sending this review
to the author in order that we may communicate directly, saving a considerable
amount of your time and mine and Laikov’s.

Sincerely yours,

Bill Hoover
Ruby Valley Nevada
Wednesday morning
\end{verbatim}
\textit{The file} ``LAIKOV.pdf'' \textit{is the article}
DOI: http://dx.doi.org/10.12921/cmst.2015.21.03.001
\\
\textit{The file} ``LAIKOV2.pdf'' \textit{is the article}
DOI: http://dx.doi.org/10.1103/PhysRevA.41.4552
\\

\noindent
\textit{Editor's message}:
\begin{verbatim}
We see that one of our referees cc'ed you on his report. We are still
seeking further review, so you should postpone any response or
resubmission until we contact you again.

We appreciate your patience in the meantime.
\end{verbatim}
\vspace{1em}
Correspondence\\
\begin{tabular}{lll}
\hline
Opened & Closed & Description \\
\hline
06Sep18&       &Editorial decision and/or referee comments sent to author \\
09Jul18&06Sep18&Review request to referee; editor concludes response unlikely \\
23Jul18&06Sep18&Review request to referee; editor concludes response unlikely \\
27Aug18&04Sep18&Review request to referee; report received \\
29Aug18&       &Status update sent to author \\
06Aug18&       &Reminder to referee [others sent (not shown) at 1-2 week intervals] \\
30Jul18&       &Reminder to referee [others sent (not shown) at 1-2 week intervals] \\
17Jul18&17Jul18&Review request to referee; message received (not a report) \\
17Jul18&       &Status update sent to author \\
29May18&12Jul18&Review request to referee; message received (not a report) \\
26Jun18&27Jun18&Review request to referee; message received (not a report) \\
26Jun18&       &Correspondence (miscellaneous) sent to author \\
20Jun18&20Jun18&Review request to referee; report received \\
29May18&20Jun18&Review request to referee; message received (not a report) \\
19Jun18&       &Reminder to referee [others sent (not shown) at 1-2 week intervals] \\
19Jun18&       &Reminder to referee [others sent (not shown) at 1-2 week intervals] \\
29May18&       &Correspondence (miscellaneous) sent to author \\
29May18&       &Right to publish signature received \\
29May18&       &Correspondence (miscellaneous) sent to author \\
21May18&       &Acknowledgment sent to author \\
21May18&       &Correspondence (miscellaneous) sent to author \\
\hline
\end{tabular}\\

\vspace{2em}
\noindent
\textit{Editor's message}:
\begin{verbatim}
The above manuscript has been reviewed by two of our referees.
Comments from the reports appear below.

These comments suggest that the present manuscript is not suitable for
publication in the Physical Review.

P.S.  We regret the delay in obtaining these reports.
\end{verbatim}
\textit{Reviewers' comments}:
\begin{verbatim}
----------------------------------------------------------------------
Report of the First Referee -- ES11735/Laikov
----------------------------------------------------------------------

I think that the paper has merit and interest as time-reversibility is
fundamental to simulation and, so far as is known, cannot be achieved
precisely in nonequilibrium simulations, which are possible using
isokinetic thermostatting equations of motion. The question which
needs to be clarified is the order of the error which can be achieved
with the proposed algorithm. The author should compare his approach to
the approach indicated in the first equation on page 4553 of LAIKOV2 (
attached above ). In order to make such a comparison a problem which
allows an isokinetic solution but which is also as simple as possible
is required. I suggest that the author solve the cell-model problem
described in PDF file (attached) and use the criteria in that paper to
describe the quality of his time reversibility. For a truly
time-reversible algorithm integer arithmetic is necessary (but not
sufficient) and this is something on which the author might wish to
comment or demonstrate.

In summary I recommend that the author include his solution to the
cell-model problem of LAIKOV. I would be happy to review the paper
with this addition and believe that it would be worth publishing in
Physical Review E in view of the interest of time-reversibility and
thermostatted equations of motion apparent in the references. I am
taking the liberty of sending this review to the author in order that
we may communicate directly, saving a considerable amount of your time
and mine and Laikov’s.

----------------------------------------------------------------------
Reference Material of the First Referee -- ES11735/Laikov
----------------------------------------------------------------------

See Attachment: es11735_refmat_1.pdf


See Attachment: es11735_refmat_2.pdf

----------------------------------------------------------------------
Report of the Second Referee -- ES11735/Laikov
----------------------------------------------------------------------

The current work proposes to use circular arc interpolation (Equation
4) for integrating the isokinetic equation of motion. A full spectrum
of numerical integrators have been well developed in the literature.
Between all these integrators the fundamental difference is the
interpolation scheme. Extra complexities may occur for simulating
physical systems, since additional important properties more than
stability and efficiency are favored, such as time reversibility and
symplecticity. The classical Verlet scheme is such a scheme and is
probably mostly applied in atomistic simulations. This present work
attempts to provide an alternative by replacing the parabola in
conventional Verlet scheme by circular arc interpolation. The subject
would be of interest to the computational physics community. However,
considering the novelty of this work, I tend to suggest the author to
submit the current work to a more suitable or specialized journal.

Still, the following are some more comments:

1. Since the author attempts to derive an alternative to the
conventional Verlet scheme, he/she should provide comparisons to
Verlet scheme in terms of numerical performance metrics, such as
efficiency and accuracy. In fact, due to expensive operations of
trigonometric functions, the efficiency the proposed scheme should be
expected to the lower than Verlet.

2. Relatedly, the author may consider motivating the current work more
carefully. On one hand, the conventional Verlet scheme offers higher
efficiency while the same accuracy, while fulfilling the requirements
of reversibility and symplecticity, why should one expect an
alternative? On the other hand, in principle, there exist infinite
number of interpolation schemes. Parabola used in Verlet could be the
simplest second-order scheme. The authors may need to justify their
choice of interpolation basis functional.

3. More than the demonstrated case, the authors may consider providing
more verification simulations. The keys of the proposed scheme are
reversibility and symplecticity. However, these important aspects are
missing in the testing cases. Moreover, the authors are suggested to
carry out more calculations and demonstrate the superiority of the
proposed scheme.
\end{verbatim}

\vspace{1em}
\noindent
11.~\textbf{Theoretical Chemistry Accounts}
\begin{tabbing}
Received: \= 09-Sep-2018 \\
Reviewed: \> 26-Sep-2018 \\
Revised:  \> 03-Oct-2018 \\
Accepted: \> 05-Oct-2018 
\end{tabbing}
\textit{Editor's message}:
\begin{verbatim}
We have now received sufficient referee advice on your manuscript:

"A reversible numerical integrator of the isokinetic equations of motion"

which you submitted to "Theoretical Chemistry Accounts".

Based on the comments of the reviewer(s), we would be pleased to reconsider
for publication a manuscript incorporating minor revisions that address
their points.  Together with preparation of your revised manuscript,
please assemble a list of responses to each point raised by the referee(s).

When you submit your revised MS, please also submit your response
to the referee(s) as a separate submission item.
\end{verbatim}
\textit{\tiny(technical details follow)}
\begin{verbatim}
We look forward to receive your revised manuscript within eight weeks.
\end{verbatim}
\textit{Reviewer's comments}:
\begin{verbatim}
Reviewer #1: This paper describes a numerical algorithm to integrate isokinetic
equations to efficiently obtain thermodynamic properties. The steps are taken
so that three consecutive points are on an arc of a circle.  A method similar
in spirit has ben used to previously to integrate the equations used to follow
steepest descent paths and should be cited
(Journal of Chemical Physics 90, 2154 (1989); https://doi.org/10.1063/1.456010,
 Journal of Chemical Physics 95, 5853 (1991); https://doi.org/10.1063/1.461606).
The second order GS method is the same as the implicit trapezoid method
described by Gear (C. W. Gear, Numerical Initial Value Problems in Ordinary
Differential Equations, Prentice-Hall, Englewood Cliffs, NJ, 1971). In addition
to test calculations that can be compared to analytical solutions, the author
uses the methodology to finds structures of water clusters and water-alanine
clusters.  Water clusters have been studied extensively. While the relative
energies of the lowest minima sensitive to the level of theory used to calculate
the energy, the topology of the lowest energy structures should be much less
sensitive.  The author should add a short paragraph comparing the lowest few
minima to the literature (for leading references, see
Journal of Chemical Physics 139, 114101 (2013); doi: 10.1063/1.4820906).

Typo: In eq 38, the LJ energy should be r^-12 - 2 r^-6
\end{verbatim}
\textit{Author's reply}:
\begin{verbatim}
Dear Editors and Reviewer,

I am glad to know that my work

"A reversible numerical integrator
of the isokinetic equations of motion"

has met with understanding and should be
of interest to the readership.

I am thankful to the Reviewer
for the helpful comments which I accept,
and now I have revised the text accordingly:

1) The last sentence in the Introduction now reads:
"
Here we design an explicit reversible integrator
of the isokinetic equations of motion
based on the local circular arc interpolation ---
the simplest analytical curve
naturally parametrized by its length,
with the simplest (constant) curvature,
fits well to the isokinetic trajectory
and plays the same role as the parabola
in the Verlet~\cite{V67} method ---
it also shares some kinship
with the well-known implicit second-order
integrator~\cite{GS89,GS91,G71}
of the first-order differential equations
of the steepest descent path.
"
All the three citations {GS89,GS91,G71}
are now included, as suggested by the Reviewer.
I must confess that I was thinking about
the connection with the Gonzalez-Schlegel (GS) method
some time ago, even though there are differences
between the two problems and methods:
the order of the differential equations
is either first or second,
the numerical integrator
is either implicit or explicit;
but they also have an important property in common:
the constant velocity norm.
I was wondering if there would be
a careful expert Reviewer
who would also note this connection,
and I am happy to see that this indeed happened.

2) The next-to-last sentence
in the Chemical Applications section now reads:
"
For the smaller systems, most if not all of the stable structures
seem to be found, and the global minimum is likely to be amoung them
(for the water clusters,
the global minima reported~\cite{WH98}
for the TIP4P~\cite{BF33,JCMIK83} empirical potential
are structurally the same
as the lowest, or seldom next-to-lowest, ones found in our tests).
"
I think that the comparison with the widely used
TIP4P force-field is meaningful,
as my electronic structure model
is also quite approximate.
For the water hexamer, the TIP4P predicts a global minimum
that switches places with the second lowest one
when calculated by a more accurate method like MP2,
my electronic structure model does also agree
with the latter, but I do not think this is worth discussion
in this work.

3) I have corrected the typo in the LJ energy,
thanks again for the careful reading!

I have also added one more citation (doi: 10.1063/1.1563597 )
that I feel is worth noting in connection
with the hybrid Monte Carlo
for which the isokinetic trajectory can also be used.

Best regards,
Dimitri Laikov
\end{verbatim}

\end{document}